\documentclass[10pt,journal,compsoc]{IEEEtran}
\newif\ifpeerreview

\peerreviewfalse

\usepackage[nocompress]{cite}
\usepackage{url}
\usepackage{amsmath,amssymb,graphicx}
\usepackage{lipsum} 
\usepackage{xcolor}
\usepackage{colortbl}
\usepackage[switch]{lineno}
\usepackage{placeins}
\usepackage{afterpage}

\definecolor{tabfirst}{rgb}{0.968,0.808,0.804} 
\definecolor{tabsecond}{rgb}{0.996,0.980,0.820} 
\definecolor{tabthird}{rgb}{0.839, 0.996, 0.816}

\newcommand{\paperID}{57}

\title{\huge Event2Audio: Event-Based Optical Vibration Sensing}

\author{Mingxuan~Cai$^{*}$, Dekel~Galor$^{*}$,  Amit~Pal~Singh~Kohli, Jacob~L.~Yates, and~Laura~Waller
\IEEEcompsocitemizethanks{\IEEEcompsocthanksitem M. Cai and D. Galor contributed equally to this work.\protect
\IEEEcompsocthanksitem M. Cai, D. Galor, A. P. Kholi, L. Waller are with the Department of Electrical Engineering and Computer Sciences, J. L. Yates is with the Herbert Wertheim School of Optometry and Vision Science, University of California, Berkeley, Berkeley, CA, 94720.\protect
\IEEEcompsocthanksitem E-mail: \{mingxuan\_cai, galor, apkohli, yates, waller\}@berkeley.edu
}
}

\begin{document}
\bstctlcite{BSTcontrol}
\IEEEtitleabstractindextext{%
\begin{abstract}
Small vibrations observed in video can unveil information beyond what is visual, such as sound and material properties. It is possible to passively record these vibrations when they are visually perceptible, or actively amplify their visual contribution with a laser beam when they are not perceptible. In this paper, we improve upon the active sensing approach by leveraging event-based cameras, which are designed to efficiently capture fast motion. We demonstrate our method experimentally by recovering audio from vibrations, even for multiple simultaneous sources, and in the presence of environmental distortions. Our approach matches the state-of-the-art reconstruction quality at much faster speeds, approaching real-time processing.

\end{abstract}

\begin{IEEEkeywords} 
Optical Vibration Sensing; Vibrometry; Event Cameras;
\end{IEEEkeywords}
}

\maketitle

\IEEEraisesectionheading{
  \section{Introduction}\label{sec:introduction}
}

\IEEEPARstart{T}{he} imperceptible vibrations in everyday objects contain rich information about their composition and the environment they interact with. Optical vibrometry enables remote measurement of such vibrations using visual sensors. This makes optical vibrometry an important tool across various engineering disciplines, where it is often used for fault detection \cite{cristalli2006mechanical, vass2008avoidance}, sound recovery \cite{wu2020specklemotion}, and inferring material properties \cite{bouman2013material, feng2022visualmaterial}.

When employing such systems for dynamic tasks like fault detection or sound recovery, vibration signals must be reconstructed quickly and reliably from optical measurements, thus creating a need for robust, high-speed optical vibrometry\cite{realtimeLDV}. In this work, we take a critical step towards this ultimate goal by developing a new vibrometer using event-based sensing \cite{gallego2020eventsurvey} that approaches real-time processing speeds while maintaining high-quality reconstruction.

Existing optical vibrometers trade off robustness, sensitivity, speed, and design simplicity. Passive vibrometers directly image the surface of a vibrating target \cite{davis2014visualaudio, bouman2013material, feng2022visualmaterial}, allowing for simple hardware at the cost of sensitivity---they can only measure motion that is visually observable \cite{sheinin2022dual}. On the other hand, active vibrometers \cite{lutzmann2011laser, zalevsky2009heartbeat} illuminate the target and measure the motion signal from reflected light, offering increased sensitivity but at the expense of robustness and simplicity \cite{cristalli2006mechanical}. Recent work uses active modalities that encode the reflected light with simple hardware but require sophisticated, slow post-processing algorithms to unearth the vibration signal \cite{zalevsky2009heartbeat, sheinin2022dual}. The key is striking a balance where the vibration is optically encoded using simple hardware while still allowing for fast decoding---something we achieve by taking advantage of the unique properties of event-based cameras.

These cameras \cite{gallego2020eventsurvey, lichtsteiner2006128eventsensor} record changes in log-intensity as asynchronous events, each carrying a timestamp, spatial location, and polarity (i.e., the sign of the intensity change). This sensing paradigm enables extremely high temporal resolution while maintaining low memory and processing overhead \cite{Prophesee2022}. The sparse and information-rich event streams that come from the event camera provide an ideal solution for co-optimizing performance and speed \cite{eventLFM}, since they reduce redundant information (i.e., static background pixels), which only serve to slow down post-processing algorithms. Inspired by their natural fit for real-time sensing, we incorporate them into a novel active vibrometry system. Our method reflects coherent light off of the target object, creating a motion-sensitive speckle image at the event sensor. The event sensor only records the aspects of the speckle that move with the object vibrations, allowing us to create a near real-time post-processing algorithm that maps events to the underlying vibration signal.

We demonstrate our method experimentally by recovering audio from vibrations, even in the presence of multiple simultaneous sources and environmental distortions. Our approach matches the state-of-the-art (SOTA) reconstruction quality at much faster speeds approaching real-time processing. We make the following contributions:
\begin{itemize}
    \item We present a novel event-based vibrometry method for recovering audio signals from vibrating surfaces. The proposed system is simple, compact, and does not require complex alignment.
    \item Leveraging this novel system, we introduce an audio reconstruction algorithm approaching real-time processing with significantly better performance than other near real-time methods.
    \item We introduce a second offline algorithm that matches SOTA performance with an order-of-magnitude speedup over existing SOTA methods.
    \item We validate our method on audio signals in varying experimental conditions, outperforming traditional microphones in challenging environments.
\end{itemize}

\begin{figure*}[!t]
\centering
\includegraphics[width=1\textwidth]{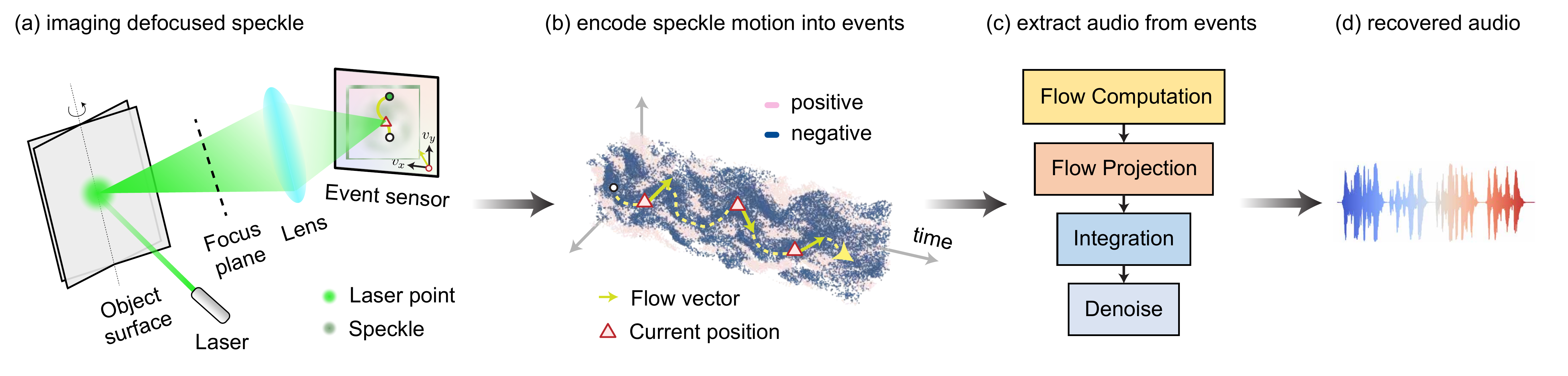}
\caption{Schematic of the proposed method. \textbf{(a)} Imaging defocused speckle. A coherent laser illuminates the vibrating surface, generating a defocused speckle pattern on the sensor plane. The pattern's 2D movements are captured by the event sensor. \textbf{(b)} The captured motion is encoded into a stream of asynchronous events. This event stream reflects the motion of the speckle pattern induced by surface vibrations. For each event, a corresponding optical flow vector, consisting of a timestamp and a 2D spatial velocity, can be extracted through optical flow computation. \textbf{(c)} Audio signal extraction from events. \textbf{(d)} Recovered audio waveform.}
\label{fig:method}
\end{figure*}
\afterpage

\section{Related Work}

\subsection{Optical Vibrometry} Optical vibrometry is generally categorized into two types: passive and active sensing. The distinction depends on whether active illumination, such as a laser, is used.

\textbf{Passive sensing}. Passive sensing methods rely on high-speed cameras to directly image vibrating surfaces \cite{buyukozturk2016smaller, davis2015visualmaterial, feng2022visualmaterial, chen2017videoinfrastructure} affected by audio sources \cite{davis2014visualaudio, howard2023event, niwa2023liveaudioevent, zalevsky2009heartbeat}. To further extract subtle vibrations with low amplitude, motion magnification algorithms are often applied \cite{liu2005motionmag, elgharib2015motionmag, feng20233motionmag}. Although passive sensing is simple to implement in hardware and does not require laser sources, it has notable limitations. For example, it struggles to capture visually imperceptible vibrations, which are associated with higher frequencies in natural signals \cite{one_over_f_audio1975}. This leads to degraded reconstruction quality for higher frequencies\cite{sheinin2022dual, howard2023event}, even if the sensor itself is fast enough. Moreover, its effectiveness significantly diminishes when applied to long-distance audio extraction \cite{bianchi2019longrange}.

\textbf{Active sensing}. Modern active sensing involves illuminating the rough surface of a vibrating object with a coherent laser source, producing a speckle pattern in the far field \cite{goodman2007speckle, alterman2021speckles}. Therefore, small tilts of the vibrating surface are translated into lateral shifts in the speckle pattern \cite{jo2015specklemotion, smith2017specklemotion, wu2020specklemotion, zizka2011specklemotion}. To enhance low-amplitude motions, the speckles are typically defocused before being recorded by the camera sensor\cite{sheinin2022dual, jo2015specklemotion, zhu2020speckledefocusmotion}. High-speed cameras are commonly used to capture the high-frequency motion of speckles \cite{davis2014visualaudio}. However, they are often cost-prohibitive and their improved temporal resolution comes at the expense of reduced spatial resolution \cite{davis2014visualaudio, sheinin2022dual}. As an alternative, line-based cameras are much cheaper but still offer comparable temporal resolution \cite{wu2020specklemotion, wu202120specklemotion, bianchi2019longrange}. They achieve this by projecting the image onto sensor rows, exposing one line at a time rather than the entire frame. Nevertheless, their reconstruction accuracy heavily depends on the alignment between the speckle motion and the sensor rows: if the motion is not parallel to the rows, the reconstructed amplitude may be significantly degraded. To address this limitation, a hybrid method combining frame-based and line-based cameras has been proposed \cite{sheinin2022dual, Zhang2023MultiSpot}. This approach leverages the speed of line-based cameras and the global capture capabilities of frame-based cameras. As a result, it enables high-fidelity reconstruction regardless of speckle motion orientation. However, this approach is complex, requiring multiple cameras, precise optical alignments, and the algorithm is still constrained by the reconstruction speed, taking hours to process a few seconds of data.

Our method implements an active sensing approach to capture imperceptible vibrations that are generally undetectable by passive sensing. Compared to previous active sensing approaches, it achieves a simple and compact optical setup without requiring multiple cameras or complex alignments. Despite this simplified design, the method delivers high-quality audio recovery comparable to state-of-the-art techniques, while maintaining much faster reconstruction speeds.

\subsection{Event-based Vision} \textit{Event cameras} \cite{mahowald1994silicon_eventcam, gallego2020eventsurvey}, also known as neuromorphic cameras, are an emerging technology for capturing fast dynamic scenes. Unlike conventional cameras that integrate light intensity over a fixed exposure time, event cameras only respond to brightness changes asynchronously and independently at each pixel at the microsecond level \cite{lichtsteiner2006128eventsensor}. The unique properties of event cameras enable unmatched capabilities for capturing high-speed motion \cite{chen2024eventbasedmotionmagnification}, making them ideal for applications such as eye tracking \cite{angelopoulos2020eventeye}, fluid particle tracking \cite{wang2020eventfluid}, and dynamic speckle analysis \cite{ge2022eventlaser}. 

\textbf{Event-based vibrometry.} Recent studies have explored the use of event cameras for vibrometry \cite{howard2023event, niwa2023liveaudioevent}. For instance, Howard  \textit{et al.} \cite{howard2023event} implemented a passive sensing setup using an event camera and recovered audio by analyzing the zero-crossings of pixels. This method enabled real-time audio reconstruction at an unprecedented speed. However, their approach relies solely on the timing of brightness changes. As a result, it produces binary waveforms that lack amplitude information, which affects the reconstruction quality. A similar passive sensing strategy was adopted by Niwa \textit{et al.} \cite{niwa2023liveaudioevent} with a different motion extraction algorithm.

Notably, both methods are implemented with a passive setting and are limited to detecting visually perceptible vibrations, such as guitar strings or blinking LEDs. These constraints significantly reduce the effective frequency range and degrade the overall reconstruction fidelity.

In contrast, our approach leverages active sensing to optically amplify subtle vibrations, enabling high-quality reconstruction over a broader frequency range. By developing an appropriate event-based processing method, we accurately track speckle motion, resulting in superior reconstruction quality and improved processing speed approaching real time.

\section{Proposed Method}
Our method consists of two main components: hardware acquisition and software processing. In the acquisition stage, we employ active illumination to encode object vibrations into the motion of a speckle pattern. An event camera is then used to efficiently capture this motion and transmit it to a computer as a stream of asynchronous events \cite{Prophesee2022}. The software stage processes these events to recover the vibration signal, which is achieved by computing and integrating the optical flow.

In the following subsections, we will discuss the details behind each component of our pipeline shown in Fig. \ref{fig:method}.

\subsection{Defocused Speckle Image Formation}
\label{SUBSEC:method-speckle}
A \textit{laser speckle} \cite{laserspecklepaper1963, goodman1975laserspeckle} is a random intensity distribution generated when a coherent laser beam reflects off a rough surface. The reflected light comprises numerous coherent wavelets, each originating from different microscopic elements of the surface. These dephased yet coherent wavelets interfere both constructively and destructively, resulting in a random spatial interference pattern known as speckle \cite{dainty2013laserspeckle}.

The laser speckle is highly sensitive to small surface changes, making it an effective tool for detecting vibrations. As illustrated in Fig. 1(a), a coherent laser forms a small spot on the rough surface of the target object. When the surface vibrates, it generates a translated speckle pattern, whose motion is subsequently magnified by the defocused imaging setup before being detected by the camera \cite{zalevsky2009heartbeat, zhu2020speckledefocusmotion}.

Although a vibrating object can exhibit multiple types of motion simultaneously, the effects of transverse and axial motion on the speckle's shape and displacement become negligible when speckle is defocused with the lens \cite{zalevsky2009heartbeat, zhu2020speckledefocusmotion}. Under these conditions, tilt motion becomes the dominant contributor to speckle displacement on the image sensor.

Moreover, the speckle shift remains consistent and global across the camera plane because the illuminated spot size is small, resulting in a large memory effect range \cite{pershin2011memory, osnabrugge2017memory}. Within this range, small tilts of the vibrating surface—such as those induced by sound waves—do not significantly alter the speckle pattern itself. Instead, the entire speckle field undergoes the same tilt, leading to a global translation on the sensor.

As a result, the camera captures a consistent global motion of the speckle pattern that directly reflects the surface tilt induced by vibration.

\subsection{Event Data}
This global translation of defocused speckle is imaged with an event camera, resulting in a stream of events (Fig. \ref{fig:method}(b)). Each event contains a timestamp $t$ in $\mu s$, 2D pixel position $(x,y)$, and polarity $p$ indicating an increase ($p=1$) or decrease ($p=-1$) in log-intensity $\log(I(x,y,t))$. A basic model of event triggering can be written as \cite{noise2image}: 

\begin{equation}
\left( \log(I(x, y, t)) - \log(I(x, y, t_0)) \right) \cdot p > \epsilon,
\end{equation}

\noindent Where $t_0$ the timestamp of the latest event in position $(x,y)$, and $\epsilon$ is the contrast threshold.

The motion of the speckle, driven by surface vibrations, is thus translated into temporal and spatial patterns within the event stream. These patterns inherently encode audio information, enabling the reconstruction of sound from the captured event stream and underlying speckle dynamics.

\begin{figure}[!t]
\centering
\includegraphics[width=\columnwidth]{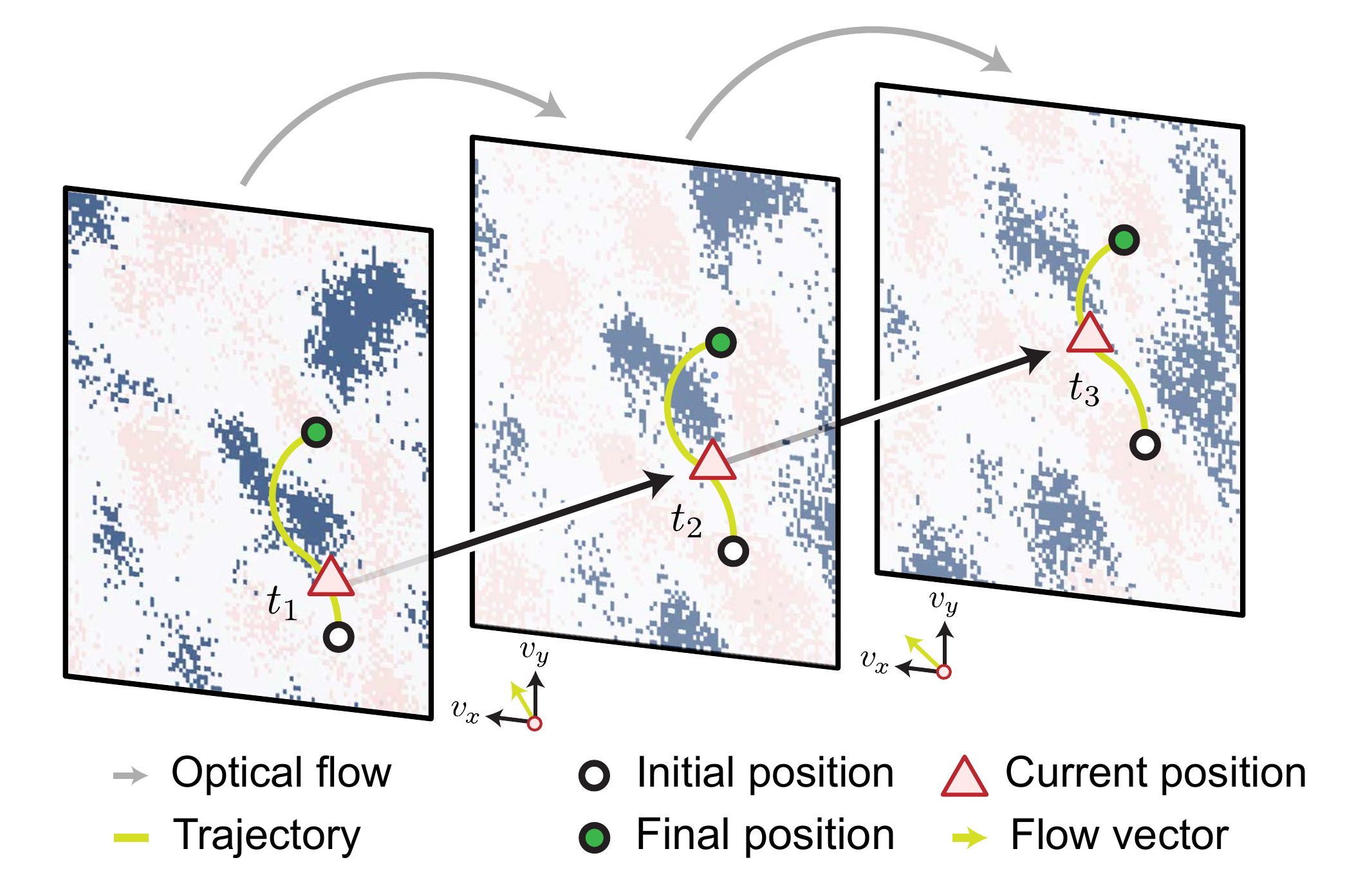}
\caption{Schematic of tracking speckle from events via optical flow. The flow indicates the current direction of motion, which can be integrated to find the motion trajectory. Pink: positive events. Blue: negative events.}
\label{fig:schematic}
\end{figure}
\subsection{Optical Flow Computation}
At this stage, we compute optical flow from the event data to estimate global motion. This is the most computationally intensive stage of our pipeline, so it must be performed efficiently to support fast and robust vibrometry reconstruction. To accommodate different applications, we provide two alternative approaches tailored for either \textit{real-time} or \textit{offline} processing, depending on the desired trade-off between reconstruction speed and quality.

The \textit{real-time} method uses a fast event-based optical flow algorithm \cite{Prophesee2022} that leverages the sparse nature of events \cite{benosman2012eventflow, ikura2024eventfeature}, operating in a streaming fashion. For each event $e_i$, characterized by its timestamp $t_i^\text{center}$, position $(x_i, y_i)$, and polarity $p_i$, the flow is estimated based on the most recent polarity-matching events in its spatial neighborhood. Specifically, for each axis, the algorithm searches for the latest events with the same polarity located $r$ pixels away in four cardinal directions: $t_i^\text{left}$ at $(x_i-r, y_i)$, $t_i^\text{top}$ at $(x_i, y_i+r)$, $t_i^\text{right}$ at $(x_i+r, y_i)$, and $t_i^\text{bottom}$ at $(x_i, y_i-r)$. We use $r=7$ pixels, which we found to perform well empirically in our experiments. The local flow is then computed as the spatial difference divided by the temporal difference. Notably, only the most recent matching event is retained per axis, allowing at most one neighboring horizontal and one neighboring vertical event to be used for flow estimation. For example, if the right pixel had a more recent timestamp than the left, the flow in the $x$ direction for event $e_i$ is computed as $\frac{\Delta x}{\Delta t}=\frac{(x_i+r)-x_i}{(t_i^\text{right}-t_i^\text{center})}$. 

Each flow estimate carries a timestamp, spatial coordinates, and a velocity vector. For global motion estimation we discard the spatial coordinates and retain only the velocity components. To obtain a temporally dense representation, we quantize timestamps to a high rate ($100$ kHz in our experiments), and aggregate all events that fall within the same quantized time bin, yielding a dense global flow signal with separate $x$ and $y$ velocity channels.

This method operates with extremely short integration windows, focusing on the temporal characteristics of motion. It offers super-fast reconstruction speed and is well-suited for applications that require live processing. However, because it ignores the broader spatial context across the object, it is more susceptible to noise and more sensitive to parameter tuning, which may impact reconstruction quality.

As an alternative, we provide an \textit{offline} implementation. Here, events are first temporally integrated to form dense frames. We then apply the Gunnar Farnebäck algorithm \cite{farneback2003two, opencv_library}, a classical optical flow method that uses a pyramidal strategy to capture motion at multiple spatial scales. The resulting flow frames are spatially averaged (each pixel weighted by the number of events it received) to get the global flow. This multiscale approach preserves broad spatial content, enabling more robust extraction of global motion while alleviating the need for extensive parameter tuning. Unlike the \textit{real-time} method, this integrate-first strategy does not leverage the asynchronous nature of events and is consequently slower. However, the integration makes the algorithm more robust to noise, and enables the use of mature image-based optical flow algorithms. Thus, the \textit{offline} method acts as a benchmark and suitable alternative when live processing is not needed. Moreover, since the hardware is identical to the \textit{real-time} method, users can seamlessly switch between the two. In practice, the \textit{offline} method is still reasonably fast and yields more stable and accurate reconstructions.

\subsection{Recovering Audio Signal from Motion}
After obtaining the 2D global motion of the speckle, we convert it into a 1D audio signal by projecting the motion traces onto a single dimension while preserving essential vibration information. In practice, we perform the projection by first temporally aligning the horizontal and vertical motion calculations to account for phase differences (using cross-correlation). The two are then averaged, and the result is integrated over time to reconstruct the 1D waveform.
Although different frequency components may oscillate along different directions, this projection enables us to retain the primary temporal characteristics of the motion necessary for accurate audio reconstruction.


\subsection{Denoising}
We further process the recovered audio signal to improve its signal-to-noise ratio. Since our reconstruction pipeline integrates motion, noise can build up over time, resulting in a gradual drift in the output signal in low frequency \cite{davis2014visualaudio}. To mitigate this effect, we apply a high-pass Butterworth filter. The cutoff frequency is carefully chosen to preserve most meaningful audio content; in our experiments, it ranged between 10 and 100 Hz.

In addition, to further reduce background noise and enhance speech intelligibility, we used the noisereduce Python library \cite{tim_sainburg_2019_3243139, sainburg2020finding}, which we found to be effective with almost negligible processing time. We empirically found that using 80\% strength, with frequency mask smoothing of 50 Hz, temporal mask smoothing of 100 ms, and a window size of 100 ms worked well across experiments and methods.


\section{Prototype and Implementation Details}
\begin{figure}[!t]
\centering
\includegraphics[width=0.4 \textwidth]{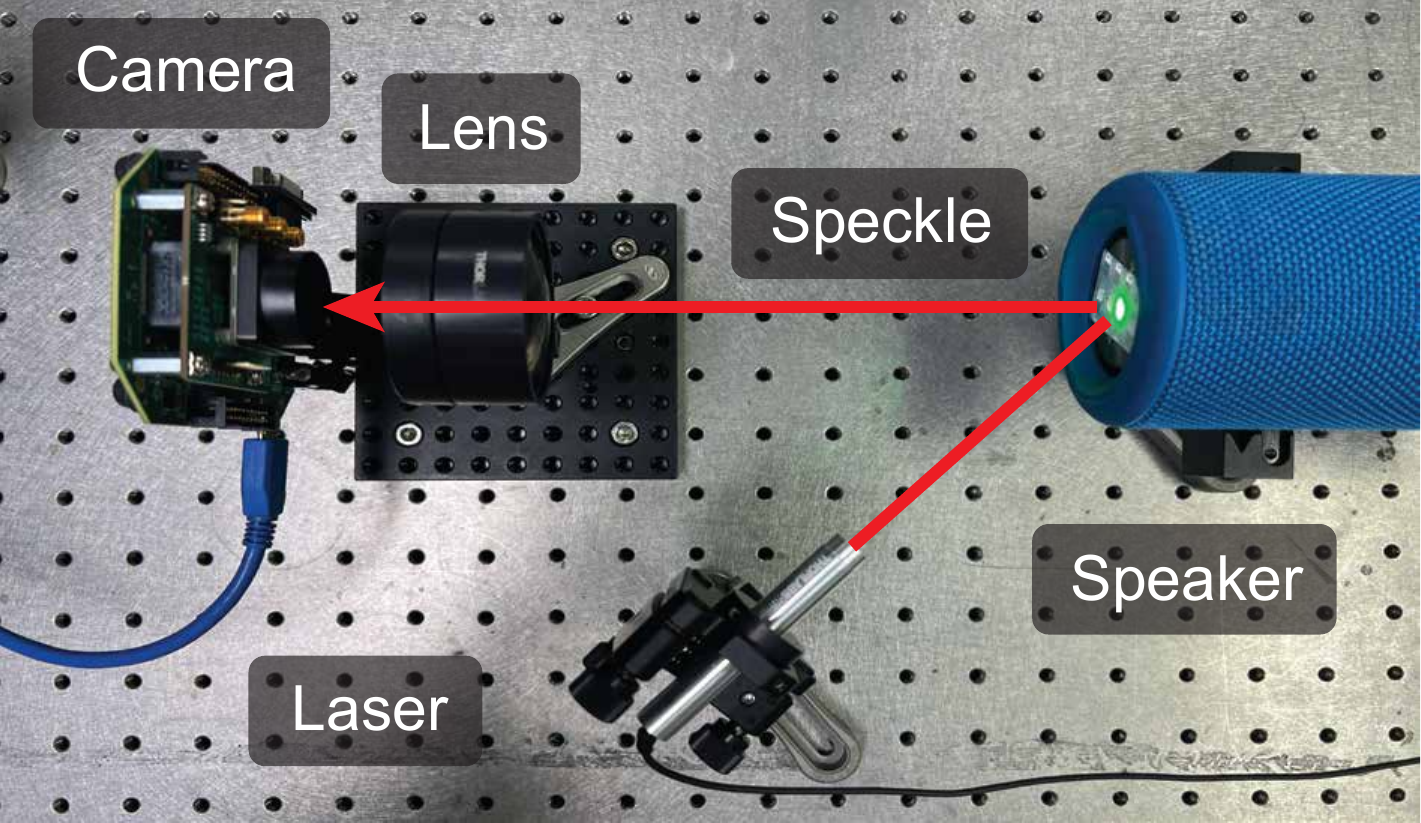}
\caption{Experimental system prototype. The laser illuminates the membrane of the speaker. Then, the reflected speckle pattern is captured by the lens and event camera. }
\label{fig:setup}
\end{figure}

\textbf{Hardware prototype.} Figure \ref{fig:setup} illustrates our prototype system. The system utilizes a 532 nm, 4.5 mW laser (Thorlabs CPS532) to illuminate the target surface, creating a small laser spot. Given the low power of the laser, a small patch of retro-reflective tape is affixed to the measured surfaces to enhance speckle intensity. Importantly, this modification does not affect the overall measurement results \cite{sheinin2022dual}. For data acquisition, the system consists of an event camera (Prophesee Metavision EVK3-HD), equipped with a 75 mm achromatic doublet lens (AC508-075-A-ML). The sensor is positioned slightly away from the focal plane of the lens to magnify the speckle pattern for subtle vibrations.

\begin{figure}[!t]
\centering
\includegraphics[width=\columnwidth]{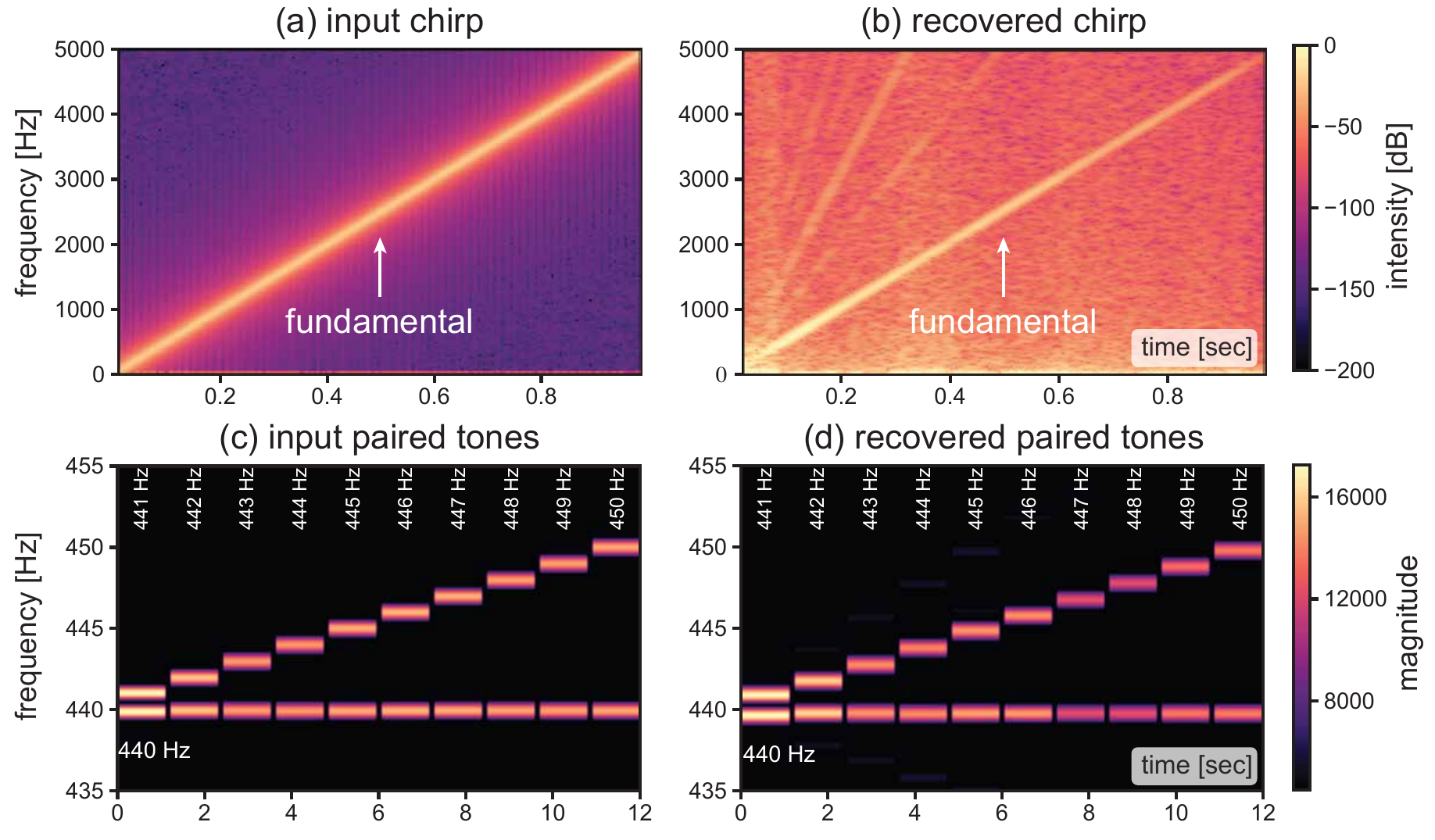}
\caption{Experimental results for recovering a chirp signal and paired tones. \textbf{(a)} A spectrogram of the input up-chirp signal sent to the speaker from 0 to 5 kHz. \textbf{(b)} Recovered chirp spectrogram using our method. \textbf{(c)} A spectrogram of the input paired tones. The reference tone is at 440 Hz with another tone from 441 Hz to 450 Hz. \textbf{(d)} Recovered paired tones spectrogram using our method. }
\label{fig:basic}
\end{figure}

\textbf{Data processing.} The default bias parameters were used for the camera, and no built-in denoising was performed on the raw event data. Processing was done on a Lenovo Legion Slim 5 Laptop with an AMD Ryzen 7 7840HS CPU. 

\textbf{Evaluation metrics.} To assess the performance of our method, we adopted evaluation metrics that are widely used in speech processing. Specifically, we employed the \textit{Perceptual Evaluation of Speech Quality} (PESQ) \cite{pesq}, \textit{Short-Time Objective Intelligibility} (STOI/intelligibility) \cite{stoi}, \textit{Mel Cepstral Distortion} (MCD) \cite{mcd}, and \textit{Log Spectral Distance} (LSD) \cite{lsd}. These metrics provide a quantitative assessment of speech quality, intelligibility, and spectral fidelity.

\section{Experimental Results}
We illustrate our method by capturing and recovering vibrations generated by various audio signals, such as chirp signals, tones, and human speech. Furthermore, we extend the utility of vibrometry to environmental distortions, including noisy, echoing, and underwater conditions, showcasing its robustness against environmental distortions and its potential for future applications. We use the built-in microphone of the iPhone 13 Pro as a reference for the octave notes and distortion experiments. Audio examples of the recovered speech are provided in the supplementary material and our website\footnote{\url{https://mingxuancai.github.io/event2audio}}. Our code and data are available on GitHub \footnote{\url{https://github.com/dgalor/Event2Audio}}.

\subsection{Capturing Audio Signals}
\subsubsection{Chirp} The most intelligible portion of human-perceived audio is typically concentrated within a frequency range up to 3.4 kHz \cite{voicefreq}. To evaluate our system’s ability to recover audio signals, we reconstructed chirp signals spanning a wide frequency range (0–5 kHz). When playing the up-chirp signal (Fig. \ref{fig:basic}(a)), the vibration of the speaker membrane induces lateral motion of the speckle on the sensor plane. As shown in Fig. \ref{fig:basic}(b), we successfully recover the fundamental frequency of the up-chirp signal. Additionally, harmonic signals emerge due to the physical properties of the speaker membrane, a phenomenon commonly observed in musical instruments such as tuning forks.
\subsubsection{Frequency Resolution for Paired Tones} Frequency resolution refers to the system's ability to differentiate between closely spaced frequencies in audio signals. To assess the frequency resolution capability of our system, we played pairs of tones with frequency separations ranging from 1 Hz to 10 Hz. As illustrated in Fig. \ref{fig:basic}(c), we set a reference tone at 440 Hz while adjusting the second tone incrementally from 441 Hz to 450 Hz. As shown in Fig. \ref{fig:basic}(d), our vibrometry approach successfully distinguished signals even with 1 Hz spacing, underscoring its ability to reconstruct complex signals with densely packed spectral components.

\begin{figure}[!t]
\centering
\includegraphics[width=\columnwidth]{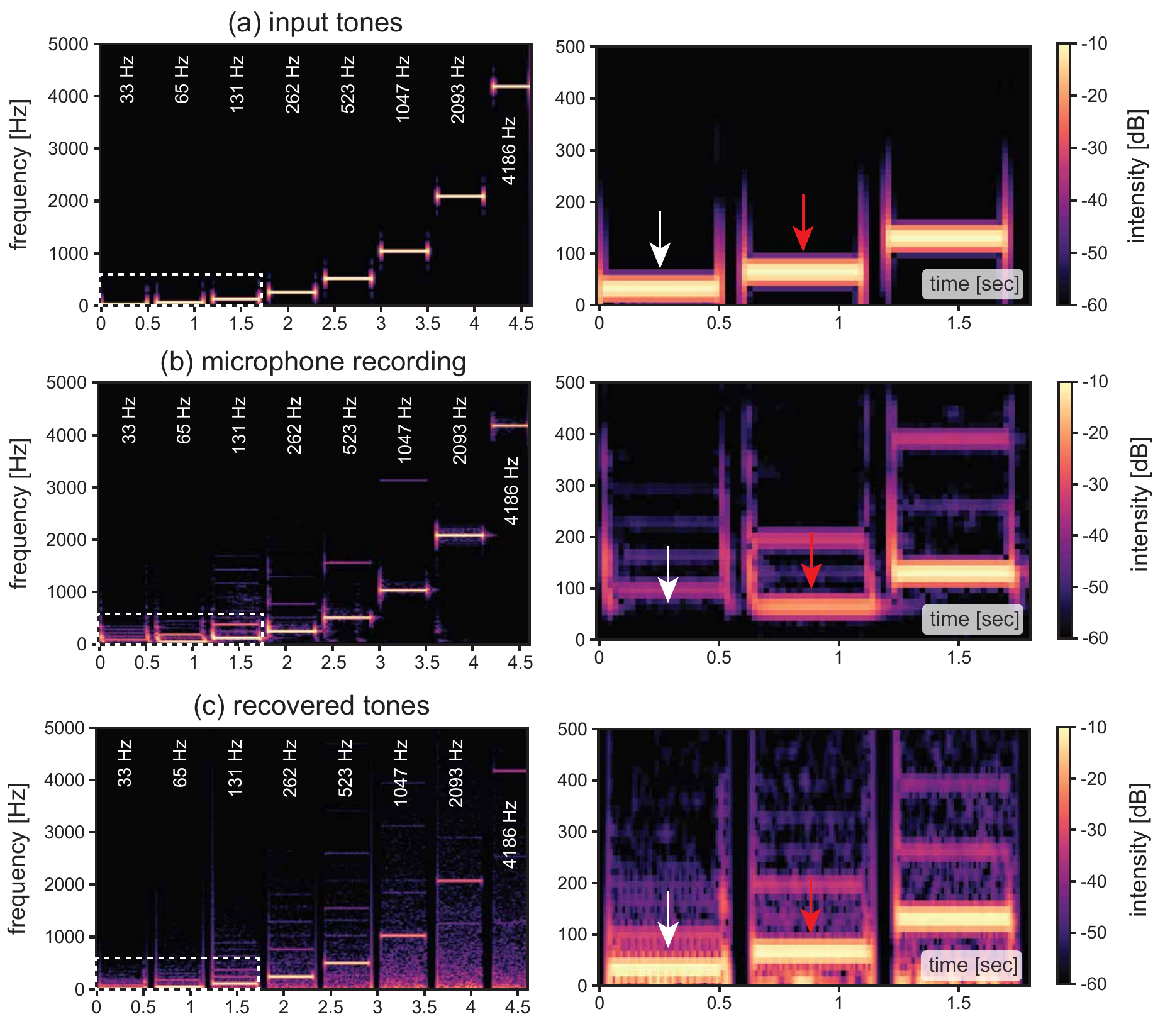}
\caption{Recovered octaves of the note C, from C1 (33 Hz) to C8 (4186 Hz). A single speaker plays eight octaves while a microphone records the audio. Our system captures the speaker membrane's vibrations and reconstructs the audio. The right column corresponds to a zoomed-in version of the left column. \textbf{(a)} Spectrogram of input tones.
\textbf{(b)} Spectrogram of tones as recorded with a microphone. It is evident that the microphone recording fails to capture low-frequency tones due to its frequency response limitations and filtering algorithms. \textbf{(c)} Spectrogram of recovered tones. By directly sensing the physical vibrations of the speaker, our system successfully recovers these tones.}
\label{fig:tones}
\end{figure}

\begin{figure*}[!h]
\centering
\includegraphics[width=1\textwidth]{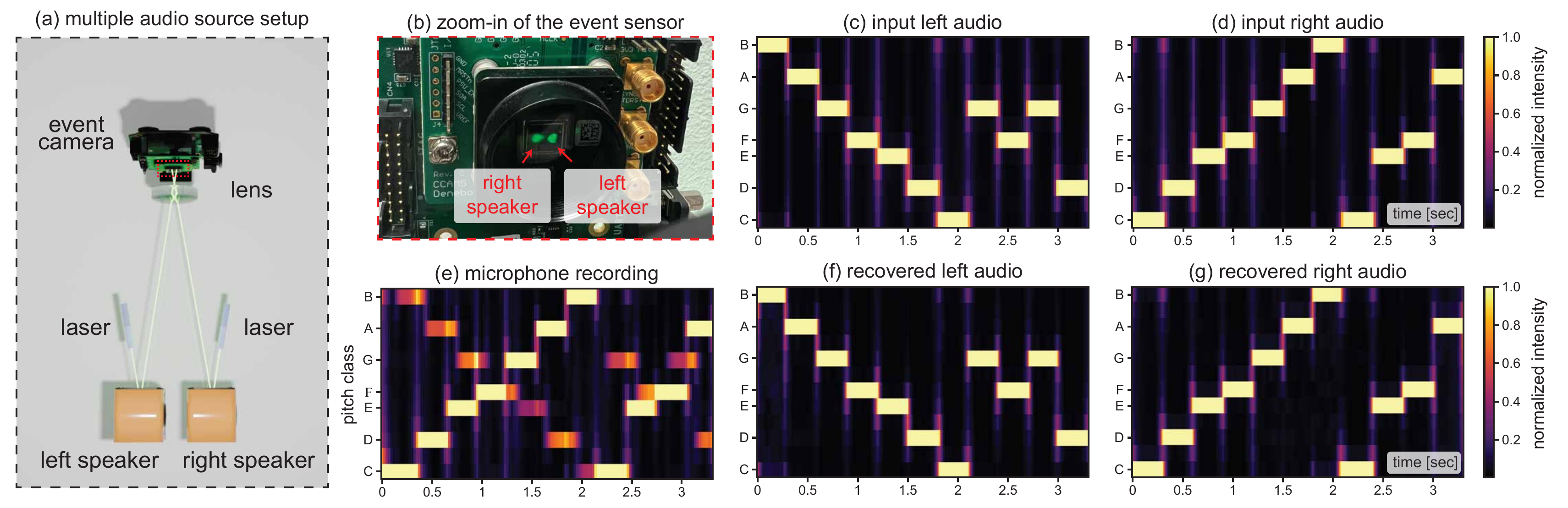}
\caption{Capturing signals from multiple audio sources. \textbf{(a)} Two lasers illuminate two different speakers simultaneously with a slightly different angle. \textbf{(b)} Zoomed-in view of the event sensor, corresponding to the red box in (a). The generated magnified speckle patterns are projected onto different parts of the event sensor. \textbf{(c-d)} The chromagrams of the input audio from left and right speakers. \textbf{(e)} The chromagram of the microphone recording. \textbf{(f-g)} The chromagrams of the recovered left and right audio signals.}
\label{fig:multiple}
\end{figure*}

\subsubsection{Note C Octaves: Vibrometry vs. Microphone} In Fig. \ref{fig:tones}(a), we played octaves of notes C1 to C8 similar to \cite{sheinin2022dual}. This spans a wide frequency range from 33 Hz to 4186 Hz. The audio was also recorded using an iPhone microphone (Fig. \ref{fig:tones}(b)) and reconstructed using our vibrometry method (Fig. \ref{fig:tones}(c)). From the result, both the microphone recording and our reconstruction exhibit high-quality performance. However, consistent with the findings in \cite{sheinin2022dual}, the microphone failed to capture the low-frequency tone at 33 Hz (white arrows in Fig. \ref{fig:tones}(right)). This is due to the low-frequency sensitivity limitations of the microphone and the built-in high-pass denoising algorithms. Interestingly, our microphone recording detected the C2 note at 65 Hz but with reduced intensity (red arrows in Fig. \ref{fig:tones}(right)), displaying a frequency response drop-off in the low-frequency region. In contrast, our vibrometry method directly measured the physical vibrations of the speaker membrane, allowing us to accurately recover both the 33 Hz and 65 Hz tones, demonstrating superior performance in low-frequency signal reconstruction.
\subsection{Remote Recording of Multiple Audio Sources}
Demixing multiple signals from different audio sources is typically a challenging task \cite{solovyev2023demix, mitsufuji2022music}. In this section, we demonstrate a configuration of our system that can simultaneously capture and separate multiple audio sources. As shown in Fig. \ref{fig:multiple}(a), two different lasers illuminate the membranes of two speakers, generating spatially separated speckle patterns on the sensor plane (Fig. \ref{fig:multiple}(b)). To assess the system performance, we played different tones for different pitch classes through two speakers (Fig. \ref{fig:multiple}(c-d)) while simultaneously recording with a microphone. We used chromagrams to analyze audio that can be meaningfully categorized into different pitch classes. Figure \ref{fig:multiple}(d) shows that the microphone captured a blended mixture of both audio signals from the left and right speakers, making it difficult to separate them directly from the recording. By leveraging the wide sensor size of the event camera, our system effectively captured vibrations from both speakers at the same time. Fig. \ref{fig:multiple}(f-g) presents the recovered chromagrams from the left and right speakers, which precisely match the input audio signals. This result highlights the effectiveness in separating multiple sound sources using our method. Additional experiments with more laser spots and audio sources are provided in the supplementary material.
\begin{figure*}[!h]
\centering
\includegraphics[width=0.8\textwidth]{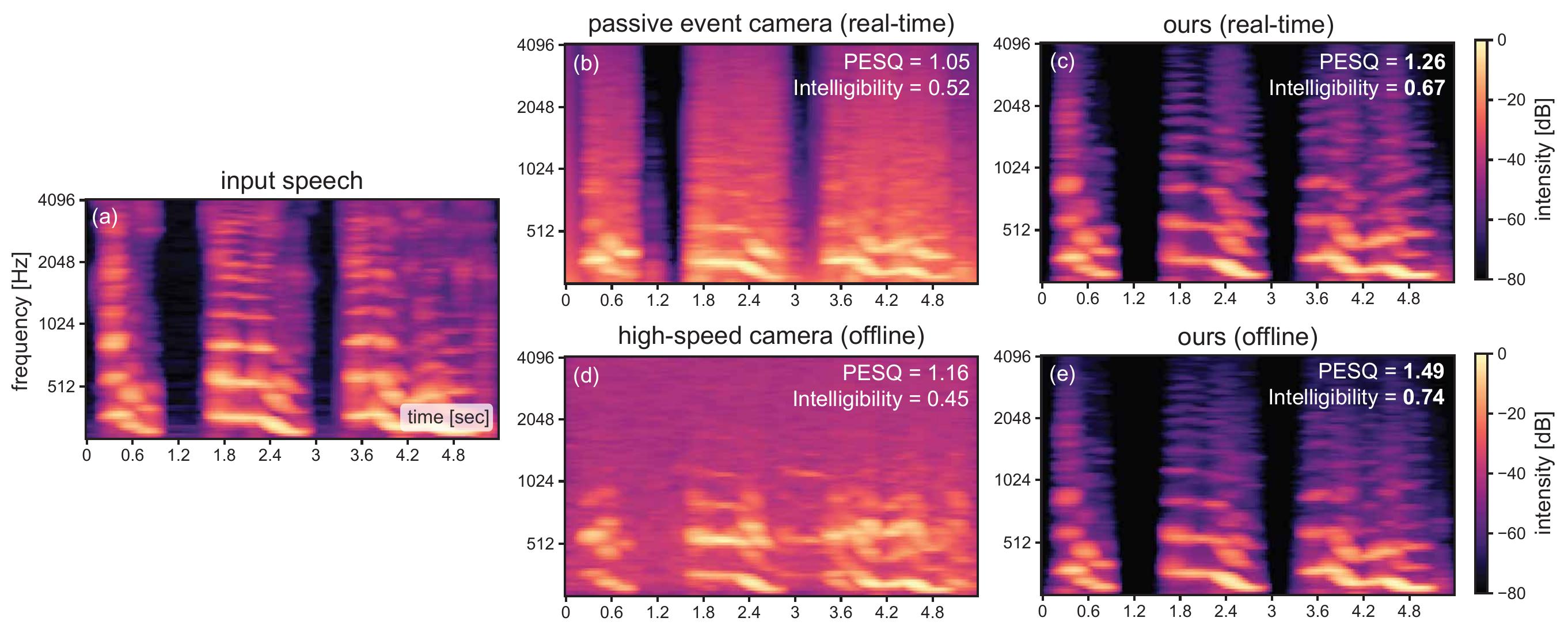}
\caption{Comparison of human speech recovery from speaker membrane with previous methods. A speaker played the Japanese audio file “Arutoki, kita…” provided by Niwa \textit{et al.} \cite{niwa2023liveaudioevent}, with baseline reconstruction results taken from that study. \textbf{(a)} Spectrogram of input speech audio. \textbf{(b)} Spectrogram of the speech audio recovered by a baseline method utilizing a passive, event-based system with real-time processing. \textbf{(c)} Spectrogram of speech audio recovered by our method with real-time processing. \textbf{(d)} Spectrogram of speech audio recovered by a method utilizing a passive, high-speed camera system with offline processing. \textbf{(e)} Spectrogram of speech audio recovered by our offline method, which provides improved reconstruction quality.}
\label{fig:japan}
\end{figure*}
\subsection{Human Speech Recovery}
\subsubsection{Speech Recovery from Speaker Membrane} \label{SUBSEC:japanese} In addition to single-frequency tones, we extend our experiment to recover human speech signals from the vibrations of a speaker membrane. Previous studies by Niwa \textit{et al.} \cite{niwa2023liveaudioevent} have achieved speech recovery from speaker vibrations but reconstruction quality still remains a bottleneck. Specifically, Niwa \textit{et al.} attached a rod to the speaker membrane, converting its vibrations into rod oscillations. However, this approach fails to accurately capture the actual vibrations of the speaker and remains ineffective in capturing high-frequency motion. Figure 6(b) and (d) illustrate the recovered Japanese speech file "Arutoki, kita..." (Fig. \ref{fig:japan}(d)), provided by Niwa \textit{et al}. While both the passive event camera and high-speed camera successfully recover frequency components up to approximately 1.1 kHz, they struggle to retain higher-frequency details, resulting in significant information loss. In contrast, in Fig. \ref{fig:japan}(c) and (d), our method achieves a more faithful and accurate recovery of high-frequency vibrations both in real-time and offline reconstruction. This is because our active sensing approach not only captures subtle membrane vibrations with high precision but also effectively amplifies them. Table \ref{table:japan} demonstrates that our method outperforms previous approaches across various evaluation metrics ([PESQ, STOI, MCD, LSD] = [1.26, 0.67, 6.72, 2.50] in real time and [1.49, 0.74, 5.57, 1.90] offline), showing a superior performance in speech reconstruction.

\begin{table}[!h]
\renewcommand{\arraystretch}{1.3}
\caption{Quantitative comparison for the experiments on recovering human speech playback from speaker membrane.}
\centering
\resizebox{\columnwidth}{!}{
\begin{tabular}{c||c|c|c|c}
\hline
\textbf{Real-time} & PESQ $\uparrow$& STOI $\uparrow$ & MCD $\downarrow$ & LSD\\
\hline
Passive event sensing \cite{niwa2023liveaudioevent} & \cellcolor{tabsecond}1.05 & \cellcolor{tabsecond}0.52 & \cellcolor{tabsecond}17.28 & \cellcolor{tabfirst}2.50\\
\hline
Ours & \cellcolor{tabfirst}1.26 & \cellcolor{tabfirst}0.67 & \cellcolor{tabfirst}5.87 & \cellcolor{tabfirst}2.50\\

\hline\hline
\textbf{Offline} & PESQ $\uparrow$& STOI $\uparrow$ & MCD $\downarrow$ & LSD $\downarrow$\\
\hline
High-speed camera \cite{niwa2023liveaudioevent} & \cellcolor{tabsecond}1.16 & \cellcolor{tabsecond}0.45 & \cellcolor{tabsecond}17.87 & \cellcolor{tabsecond}3.50\\
\hline
Ours & \cellcolor{tabfirst}1.49 & \cellcolor{tabfirst}0.74 & \cellcolor{tabfirst}5.57 & \cellcolor{tabfirst}1.90\\
\hline
\end{tabular}
}
\label{table:japan}
\end{table}

\subsubsection{Speech Recovery from a Chip Bag} 
\label{SUBSEC:chipbag}
\begin{figure*}[!t]
\centering
\includegraphics[width=1\textwidth]{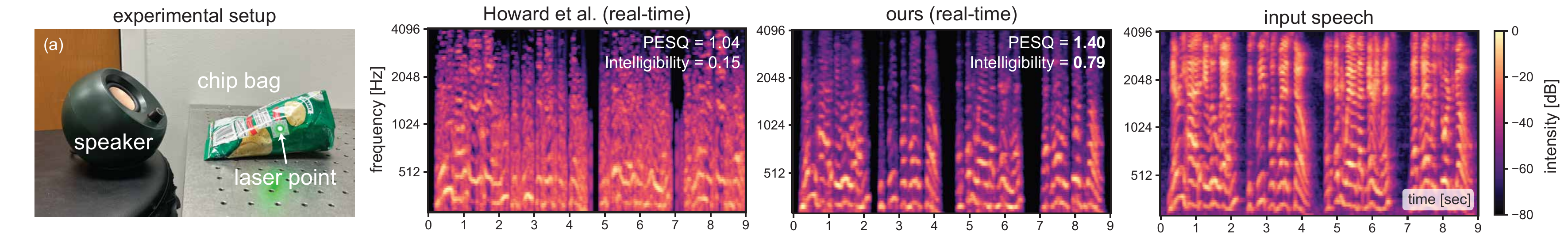}
\caption{Comparison of human speech recovery from chip bag with previous methods for real-time reconstruction. \textbf{(a)} We replicate the chip bag experiment originally proposed by Davis et al. \cite{davis2014visualaudio}. In our setup, a speaker plays the "Mary Had a Little Lamb..." audio provided by Davis et al., inducing vibrations in the chip bag. These vibrations then move the speckle pattern, which is captured by our system. \textbf{(b)} Recovered audio from Howard et al. \cite{howard2023event}, which used a zero-crossing algorithm with an event camera for passive sensing. \textbf{(c)} Recovered audio using our method. \textbf{(d)} Ground truth speech audio.}
\label{fig:chipbag1}
\end{figure*}

In Fig. \ref{fig:chipbag1}(a), we replicate the chip bag experiment originally proposed by Davis \textit{et al.} \cite{davis2014visualaudio}. In our setup, we directed a laser onto the surface of a chip bag while playing the audio "Mary Had a Little Lamb...", as provided by Davis \textit{et al}. In Fig. \ref{fig:chipbag1}(b-c), we compared our real-time reconstruction results with a previous event-based vibrometry method proposed by Howard \textit{et al.} \cite{howard2023event}. Their approach relies on detecting zero-crossings at individual pixels and can only recover a binary amplitude waveform, resulting in degraded reconstruction quality. In contrast, our method successfully recovers the audio with high quality, which outperforms Howard's approach in all evaluation metrics in Table \ref{table:chipbag}. The scores are [PESQ, STOI, MCD, LSD] = [1.04, 0.15, 26.10, 3.26] and [1.40, 0.79, 14.13, 2.41], ours being the latter.

\begin{figure*}[!t]
\centering
\includegraphics[width=1\textwidth]{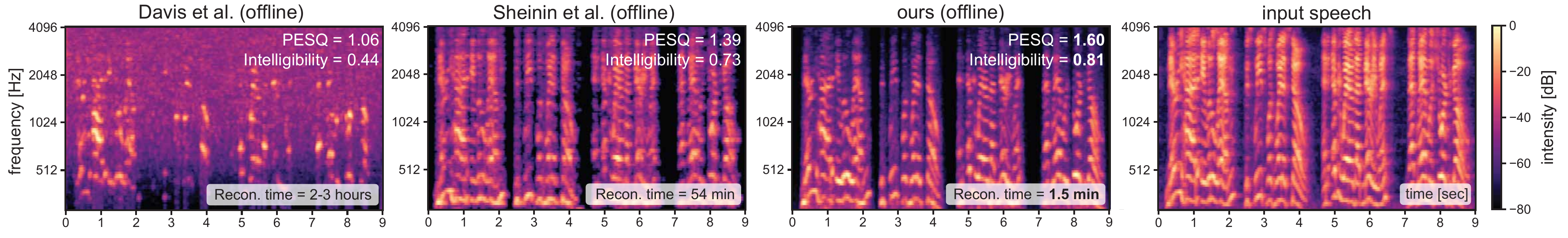}
\caption{Comparison of human speech recovery from chip bag with previous offline methods. \textbf{(a)} Recovered audio from Davis et al. \cite{davis2014visualaudio}, which employed a high-speed camera along with a strong light source. The reconstruction generally takes 2-3 hours. \textbf{(b)} Recovered audio from Sheinin et al. \cite{sheinin2022dual}, which utilized a cylindrical lens in combination with global shutter and rolling shutter cameras. The reconstruction takes approximately 54 minutes. \textbf{(e)} Recovered audio using our method. The reconstruction only takes 1.5 minutes. \textbf{(b)} Ground truth speech audio.}
\label{fig:chipbag2}
\end{figure*}

Furthermore, other vibrometry approaches \cite{davis2014visualaudio, sheinin2022dual} that achieve high-quality reconstruction typically depend on computationally intensive audio recovery algorithms, often requiring several hours to reconstruct only a few seconds of audio. In Fig. \ref{fig:chipbag2}, we compare the reconstruction quality and processing time of our offline method with prior offline approaches. Despite different setups, which makes a direct and fair comparison difficult, it is still clear that our system demonstrates comparable reconstruction quality to the state-of-the-art method proposed by Sheinin \textit{et al.}\cite{sheinin2022dual}, as shown in Table \ref{table:chipbag}, while significantly reducing reconstruction time by at least 30$\times$—from approximately 54 minutes to just 1.5 minutes.

More importantly, Sheinin \textit{et al.} employed a complex system that combines global shutter and rolling shutter cameras, along with a cylindrical lens to project speckle patterns onto the rolling shutter sensor columns. This design introduces significant challenges in optical alignment and increases hardware complexity. Additionally, splitting the optical path between two cameras reduces light efficiency, further limiting its practicality for real-world applications.

In summary, our method features simple and compact optics, eliminating the need for complex optical alignment. Additionally, it leverages the efficient data processing capabilities of event-based sensing, enabling high-quality reconstruction with significantly faster processing on the order of minutes up to real time.


\begin{table}[!h]
\renewcommand{\arraystretch}{1.3}
\caption{Quantitative comparison for the experiments on recovering human speech playback from chip bag.}
\centering
\resizebox{\columnwidth}{!}{
    \begin{tabular}{c||c|c|c|c|c}
    \hline
     \textbf{Real-time} & PESQ $\uparrow$ & STOI $\uparrow$ & MCD $\downarrow$ & LSD $\downarrow$ & Time $\downarrow$\\
     \hline
    Howard \cite{howard2023event} &  \cellcolor{tabsecond}1.04 &  \cellcolor{tabsecond}0.15 &  \cellcolor{tabsecond}26.10 & \cellcolor{tabsecond}3.26 & \cellcolor{tabfirst}real-time\\
    \hline
    Ours
    & \cellcolor{tabfirst}1.40 & \cellcolor{tabfirst}0.79 & \cellcolor{tabfirst}14.13 & \cellcolor{tabfirst}2.41 & \cellcolor{tabfirst}real-time\\
    \hline\hline
    \textbf{Offline} & PESQ $\uparrow$ & STOI $\uparrow$ & MCD $\downarrow$ & LSD $\downarrow$ & Time $\downarrow$\\
    \hline
    Davis \cite{davis2014visualaudio} & 1.06 & 0.44 & 18.31 & \cellcolor{tabsecond}2.16 & 2-3 hours\\
    \hline
    Sheinin \cite{sheinin2022dual} & \cellcolor{tabsecond}1.39 & \cellcolor{tabsecond}0.73 & \cellcolor{tabsecond}12.79 & 2.20 & \cellcolor{tabsecond}54 min\\
    \hline
    Ours & \cellcolor{tabfirst}1.60 & \cellcolor{tabfirst}0.81 & \cellcolor{tabfirst}12.56 & \cellcolor{tabfirst}1.72& \cellcolor{tabfirst}1.5 min\\
    \hline
    \end{tabular}
}
\label{table:chipbag}
\end{table}


\subsection{Vibrometry Against Environmental Distortion}

A microphone captures all sounds within its range, known as ambient sound capture, but this can introduce environmental distortions such as background noise, echoing in enclosed spaces, or attenuation underwater due to the sound waves interacting with the environment. In this section, we first propose three compelling applications where vibrometry outperforms a traditional microphone by mitigating environmental distortions. These examples demonstrate the advantages of vibrometry and highlight its potential for future applications beyond conventional audio sensing.
\subsubsection{Demix Audio from Noisy Scenes} Imagine you are in a conference room, but the speaker's voice is drowned out by background noise from the audience. In such cases, you may seek a method to eliminate background noise. Noise removal from audio has been extensively studied for decades \cite{yu2008denoise, purwins2019deepdenoise, xie2021bioacousticdenoise}. Existing approaches typically post-process the mixed audio signal via band filtering and, more recently, machine learning. These approaches are flexible in separating various, and potentially many, different sources from each other---even ones that were co-located during recording. However, their quality is largely limited by the difficulty of the problem and they often require massive paired datasets to achieve even moderate performance \cite{liu2024separate, wang2025soloaudio}. In contrast, vibrometry allows for the physical isolation of the target audio source from its noisy environment at the time of recording. This means each source is essentially recorded independently, allowing for much higher quality source separation. For example, in our work, the system actively senses the surface vibrations of the target speaker, enabling direct access to its sound signal before it propagates and mixes with other sources in the environment. This contrasts with microphones, which capture a composite sound field containing both the target and background noise. However, this does come at the cost of flexibility, since the sources in question must be physically separated and there must be sufficient space on the sensor to adequately disambiguate the signal from each source.

To evaluate our approach, we simulate a noisy environment where two speakers are generating audio; one is our target speaker, and the other speaker's audio is considered background noise. As shown in Fig. \ref{fig:distortion}(top), the target speaker is illuminated by our laser. This setup enables us to actively isolate the vibration source, eliminating distortions introduced by environmental noise. In this experiment, the target speaker plays the audio "The law of the school..." (Fig. \ref{fig:distortion}(c)), while the noise source simultaneously plays a song, creating a challenging audio separation scenario. The spectrogram of the microphone recording in Fig. \ref{fig:distortion}(a) reveals that the target audio is heavily masked by background noise, resulting in a poor PESQ score of 1.06.

In contrast, the spectrogram of our recovered speech (Fig.~\ref{fig:distortion}(b)) demonstrates significant improvement. Because vibrometry directly captures vibrations from the target source—remaining unaffected by ambient noise—we are able to extract a much cleaner speech signal. As a result, overall speech clarity is substantially enhanced, achieving a PESQ score of 1.27 and an intelligibility score of 0.80.

\subsubsection{Vibrometry Eliminates Echoes} Echoes \cite{ricker2012echo} occur in enclosed spaces where sound waves reflect off surfaces, creating overlapping signals that distort audio clarity. This effect is particularly common in large conference rooms, open offices, and empty halls. As a result, reducing echoes can significantly enhance speech intelligibility, leading to improved communication quality. 

Previous approaches \cite{sondhi1967adaptiveecho, mader2000echostep, paleologu2011sparse} for echo elimination primarily relied on post-processing recorded audio using sophisticated echo cancellation algorithms. These methods are often computationally intensive and may struggle to isolate the original signal from complex, overlapping reflections—particularly in highly reverberant environments or when echo characteristics vary dynamically \cite{benesty2001echo}. To overcome these limitations, we propose a vibrometry-based approach that eliminates the effects of echo by directly sensing surface vibrations, thereby bypassing the acoustic interference present in the surrounding environment.

In Fig. \ref{fig:distortion}(middle), we present our experimental setup inside an empty room, where a speaker plays the audio "A laudable regard..." (Fig. \ref{fig:distortion}(f)). As sound waves are emitted, they reflect off the walls, producing echoes in the form of delayed and overlapping versions of the original sound. 

To mitigate the impact of echoes, we directed a laser at the target speaker to measure its physical vibrations, enabling the recovery of speech signals to be unaffected by echo-induced distortions. Figure \ref{fig:distortion}(e) demonstrates fine spectral structure closely matching the input speech signal, whereas the microphone recording (Fig. \ref{fig:distortion}(d)) exhibits spectral smearing along the time axis due to delayed reflections of echoes. Additionally, the repeated echoes gradually decrease in intensity, forming a fading trail that further degrades audio quality.

Since it is difficult to acquire an ideal echo chamber, echoes preserved most spectral features and did not severely distort phonetic content in the recording. Therefore, intelligibility (STOI) remains relatively high (0.73) for the echoed recording with our recovery in 0.81. Besides, this metric primarily evaluates short-time intelligibility and spectral similarity, which are less impacted by moderate echoes. In contrast, PESQ, which evaluates long-term perceptual quality, penalizes echo artifacts, resulting in a greater improvement in our method (1.64) compared to microphone recordings (1.29). To further quantify this effect and fully demonstrate the advantage of vibrometry in echo-prone environments, future work will conduct experiments in a controlled echo chamber to refine and validate the system’s performance.

\subsubsection{Underwater Vibrometry} Detecting underwater sound is crucial for various scientific and industrial applications (e.g. underwater communication \cite{bass2003physical} and marine biology monitoring \cite{hawkins1986underwaterfish}). However, achieving high-quality underwater audio recovery remains a significant challenge due to strong signal attenuation, limited detectable bandwidth, and the multi-path propagation of sound waves. Given that vibrations remain preserved underwater, vibrometry appears to be a promising alternative for underwater audio recovery.

\begin{figure*}[!t]
\centering
\includegraphics[width=1\textwidth]{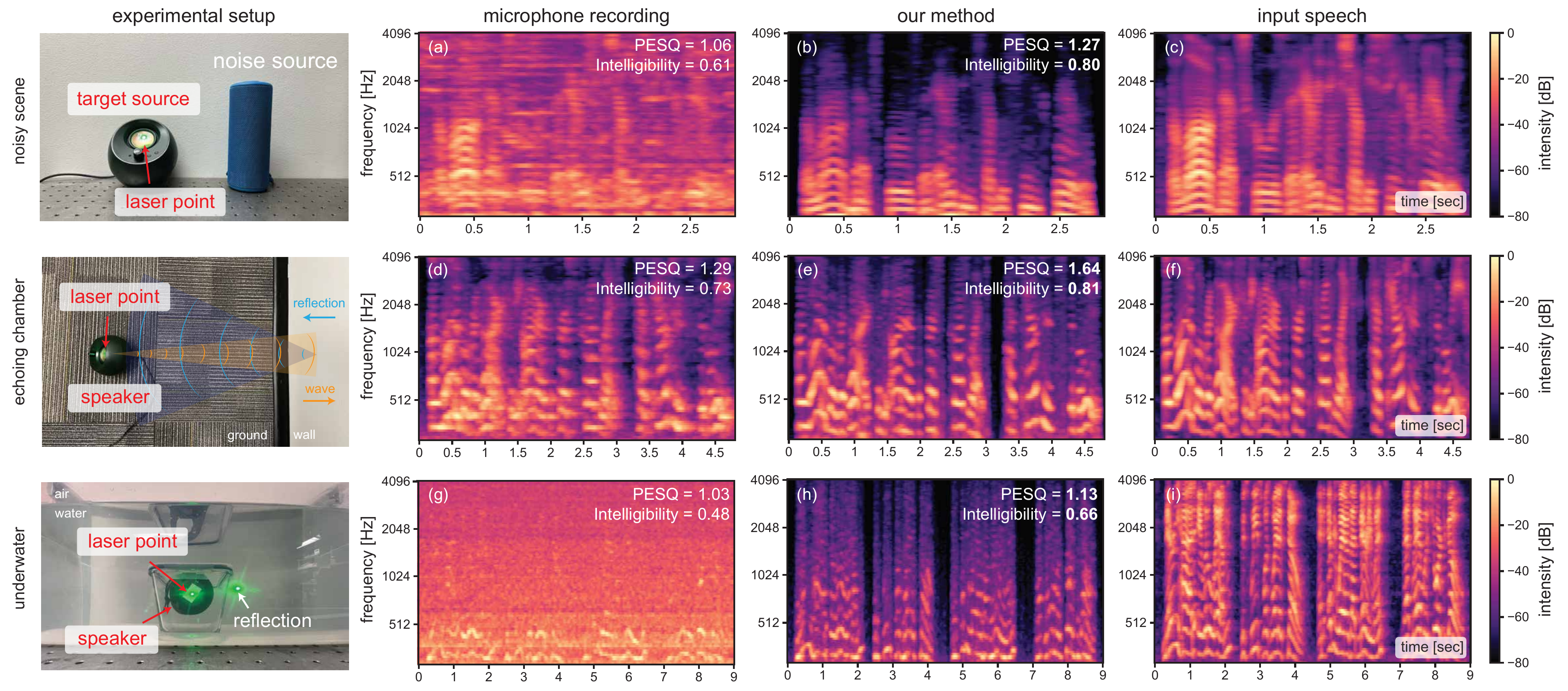}
\caption{Vibrometry against environmental distortion. \textbf{(a-c)} We construct a noisy environment where the target speaker is influenced by environmental sources. To isolate the target audio signal, we direct a laser onto the audio source playing the speech "The law of the school...", while another speaker plays interfering background noise. The microphone records a mixture of the target speech and noise, whereas the vibrometry measurement successfully isolates the speech from the noisy scene. \textbf{(d-f)} We position the system in an empty room to simulate an echo chamber with the speaker playing the audio "A laudable regard...", where reverberation degrades the audio quality. The echo introduces a "delay" effect in the microphone recording, reducing audio resolution in the spectrogram. In contrast, the vibrometry reconstruction preserves the fine details of the spectrogram. \textbf{(g-i)} We place a waterproof speaker playing "Marry had a little lamb..." inside a transparent box filled with water, which significantly attenuates the energy of the audio signal, allowing only a small fraction to be captured by the microphone. However, using vibrometry we successfully recovered the speech signal despite the distortion caused by the water.}
\label{fig:distortion}
\end{figure*}

Existing underwater vibrometry methods \cite{harland2007underwatervib, harland2004nonperturbing} are either costly or require physical contact, leading to challenges in installation complexity and material compatibility. Moreover, effective vibrometry techniques for recovering complex underwater audio signals, such as human speech, have yet to be developed.

We conducted a proof-of-concept experiment to demonstrate the feasibility of underwater audio recovery using our method. As shown in Fig. \ref{fig:distortion}(bottom), we positioned our system on a table and placed a waterproof speaker inside a sealed transparent box filled with water. To prevent it from floating, we secured the speaker inside a smaller water-filled box, anchoring it at the bottom of the enclosure. To capture the vibrations, we directed a laser onto the speaker membrane. While some laser light was reflected by the box surface, creating an additional laser point, this reflection did not align with the speckle pattern originating from the speaker membrane and therefore had a negligible impact on the measurement process.

Although underwater factors such as optical absorption, turbulence, and other unpredictable conditions can influence performance, we assume light reflection loss and water resistance on the speaker membrane to be dominant in this proof-of-concept experiment. More discussion of underwater distortions is provided in the Discussion section.

In Fig. \ref{fig:distortion}(g), the microphone recording captures primarily low-frequency signals, as most high-frequency components are lost after propagating through the water ([PESQ, Intelligibility] = [1.03, 0.48]). In contrast, our method (Fig. \ref{fig:distortion}(h)), which directly measures the physical vibrations of the speaker membrane, achieves improved audio reconstruction ([PESQ, Intelligibility] = [1.13, 0.66]). Although the laser signal also experiences absorption, attenuation, and minor turbulence caused by the water, it provides superior reconstruction quality compared to microphone recordings.

\section{Discussion}

\textbf{Light efficiency:} The amount of light captured by the sensor significantly impacts the performance of our method. Similar to previous work \cite{sheinin2022dual}, we used retro-reflective markers to enhance the speckle signal. However, unlike their approach, our method does not require light splitting, effectively doubling the number of photons available for detection. Additionally, while higher-power lasers \cite{bianchi2019longrange} can further enhance performance, the inherently high dynamic range of event sensors \cite{gallego2020eventsurvey} (140 dB vs. 60 dB in standard cameras) reduces the dependence on laser power, offering a more adaptable sensing approach. \\
\textbf{Handling large object motion:} Subtle vibrations are often superimposed with large-scale object or scene motion when capturing vibrations from musical instruments \cite{sheinin2022dual} or hand-held objects. Event cameras are well-suited for capturing both subtle vibrations and large motions simultaneously. Additional experiments that evaluate our system under substantial object motion are provided in the supplementary material.\\ 
\textbf{Event sensitivity:} The sensitivity of the event sensor impacts its performance across different levels of motion and can be adjusted through bias settings. In our system, we used the default bias setting for all experiments. Careful tuning of these parameters could enhance the signal-to-noise ratio, improve sensitivity to speckle motion, and ultimately lead to higher-quality audio reconstruction.\\ 
\textbf{Motion dynamic range:} Different from traditional cameras, the motion dynamic range of an event camera is influenced by its contrast sensitivity threshold. This dependence can present challenges when attempting to detect subtle motion below the threshold. In our experiments, system performance degrades as the magnitude of subtle motion decreases. Additional results that quantify performance over a range of motion amplitudes are provided in the supplementary material.\\ 
\textbf{Underwater distortion:} We performed a proof-of-concept experiment to recover audio signals underwater. However, distortions in real underwater environments could be far more severe. Factors such as turbulence and light scattering, which were negligible compared to the signal in our experiment, may significantly affect the final measurement. To address these challenges, potential solutions could be adapted from dynamic scattering correction techniques \cite{gigan2022wavefront, ruan2020dynamic, yi2022dynamic}. \\
\textbf{Event-based flow processing:} We explored one possible event flow algorithm for real-time audio recovery. This approach allowed for real-time reconstruction speed while achieving high-quality reconstruction. However, the algorithm remains sensitive to noise and requires hyperparameter tuning. Nevertheless, event-based optical flow tracking algorithms are advancing rapidly \cite{gallego2020eventsurvey, benosman2012eventflow, shiba2022eventflow}, and as these methods continue to improve, our event-based vibrometry approach is expected to become more robust and faster to run. \\
\textbf{Projecting global 2D motion to 1D:} Since different frequencies may oscillate in different directions, there could be improvements to the projection technique by operating in the Fourier domain and having a frequency-adaptive projection method. If we assume that there is one oscillation direction at any given time, it may still be beneficial to use a time-adaptive projection direction (perhaps incremental PCA with a forgetting factor \cite{iPCA}). We found that in practice, projecting on the first (non-time-adaptive) principal component matched the performance of our current method, so we opted for the simpler approach of averaging. \\
\textbf{Audio denoising and enhancement:} Recovering audio from vibrations inherently introduces noise into the signal \cite{davis2014visualaudio}. Additionally, materials naturally act as low-pass filters, attenuating high-frequency audio components \cite{bouman2013material}. Thus, denoising and enhancement post-processing are essential for improving audio quality. However, most learning-based methods \cite{purwins2019deepdenoise} are trained on conventional noisy audio datasets, which have different noise characteristics than vibration-induced noise. Future work should focus on training denoising and enhancement models on vibration-derived audio, optimizing both performance and reconstruction quality.\\ 

\section{Conclusion}
We propose a novel high-speed optical vibrometry approach using an event camera. Our system is simple and compact, featuring a fast audio recovery algorithm. We demonstrate its effectiveness across various audio sources, including human speech, in different environments. Our experiments showed additional evidence that optical vibrometry can successfully capture audio in difficult scenarios, like underwater, where traditional microphones suffer. We hope to see future work continue to improve the speed and robustness of vibrometry for live sensing of imperceptible signals.

\ifpeerreview \else
\section*{Acknowledgments}
The authors would like to thank Leyla Kabuli, Ethan Weber, Mark Sheinin, Matan Kichler, Rev. zFsZ, Rev. C4Q6, and Rev. uFwf for fruitful discussion, Ruizhi Cao and Tingle Li for manuscript review. This work was supported by STROBE: A National Science Foundation Science \& Technology Center under Grant No. DMR 154892 and Weill Neurohub Investigators Program Holographic all-optical electrophysiology: A new platform for ultra-fast bidirectional brain machine interfaces. This material is also based upon work supported by the National Science Foundation Graduate Research Fellowship Program under Grant No.~DGE-1752814 (DG). Any opinions, findings, conclusions, or recommendations expressed in this material are those of the author(s) and do not necessarily reflect the views of the National Science Foundation. Additionally, DG was funded by the Center for Innovation in Vision and Optics. The authors thank the developers of the software packages used in this project and not mentioned in the main text, including PyTorch \cite{paszke2019pytorch}, NumPy \cite{harris2020numpy}, SciPy \cite{SciPy2020}, matplotlib \cite{hunter2007matplotlib}, pesq \cite{pesq} and librosa \cite{librosa}. Figure style inspired by Chugunov et al. \cite{chugunov2024neural}.
\fi

\bibliographystyle{IEEEtran}
\bibliography{references}

\ifpeerreview \else

\begin{IEEEbiography}[{\includegraphics[width=1in,height=1in,clip,keepaspectratio]{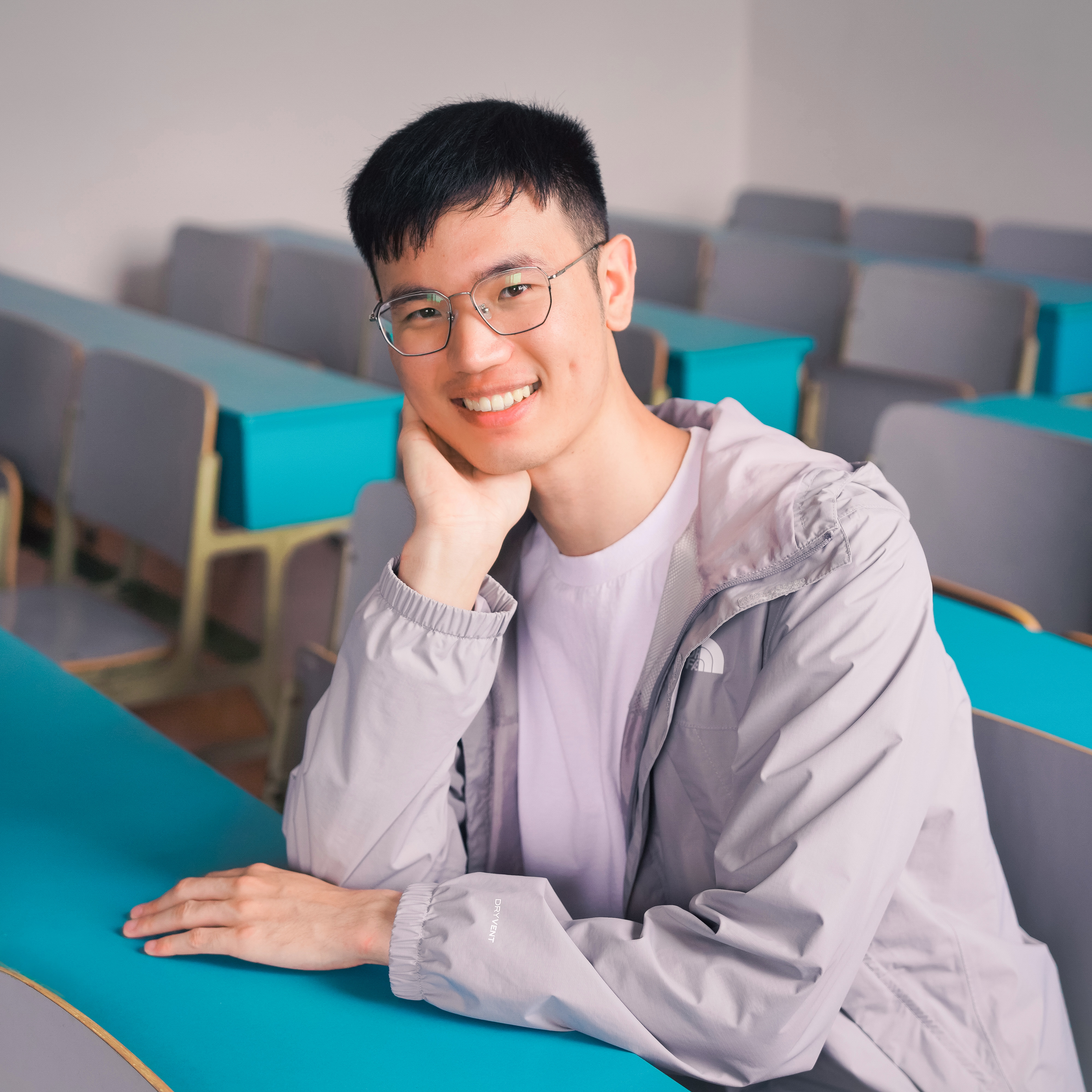}}]{Mingxuan Cai}
is a Ph.D. student in Electrical Engineering and Computer Sciences at the University of California, Berkeley, advised by Prof. Laura Waller. He received the B.S. degree in Optical Engineering from Zhejiang University, Hangzhou, China, in 2022. His current research interest focuses on space-time imaging, event-based vision, and computational fluorescence imaging.
\end{IEEEbiography}

\begin{IEEEbiography}[{\includegraphics[width=1in,height=1in,clip,keepaspectratio]{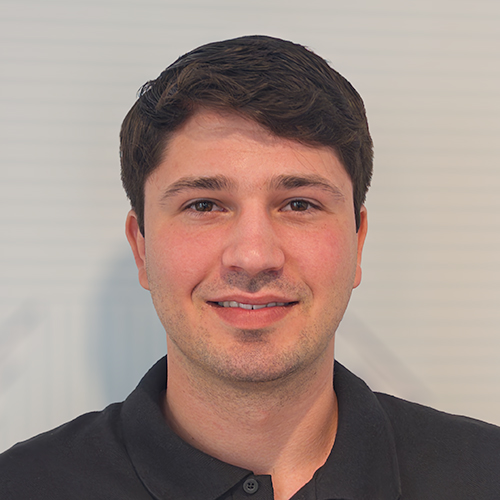}}]{Dekel Galor} is a Ph.D. student in Electrical Engineering and Computer Sciences at the University of California, Berkeley, where he is co-advised by Prof. Laura Waller and Prof. Jacob Yates. He received the B.S. degree in Electrical Engineering and Computer Sciences from UC Berkeley in 2022. His research lies at the intersection of computational imaging, visual neuroscience, and theoretical machine learning.

\end{IEEEbiography}

\begin{IEEEbiography}[{\includegraphics[width=1in,height=1in,clip,keepaspectratio]{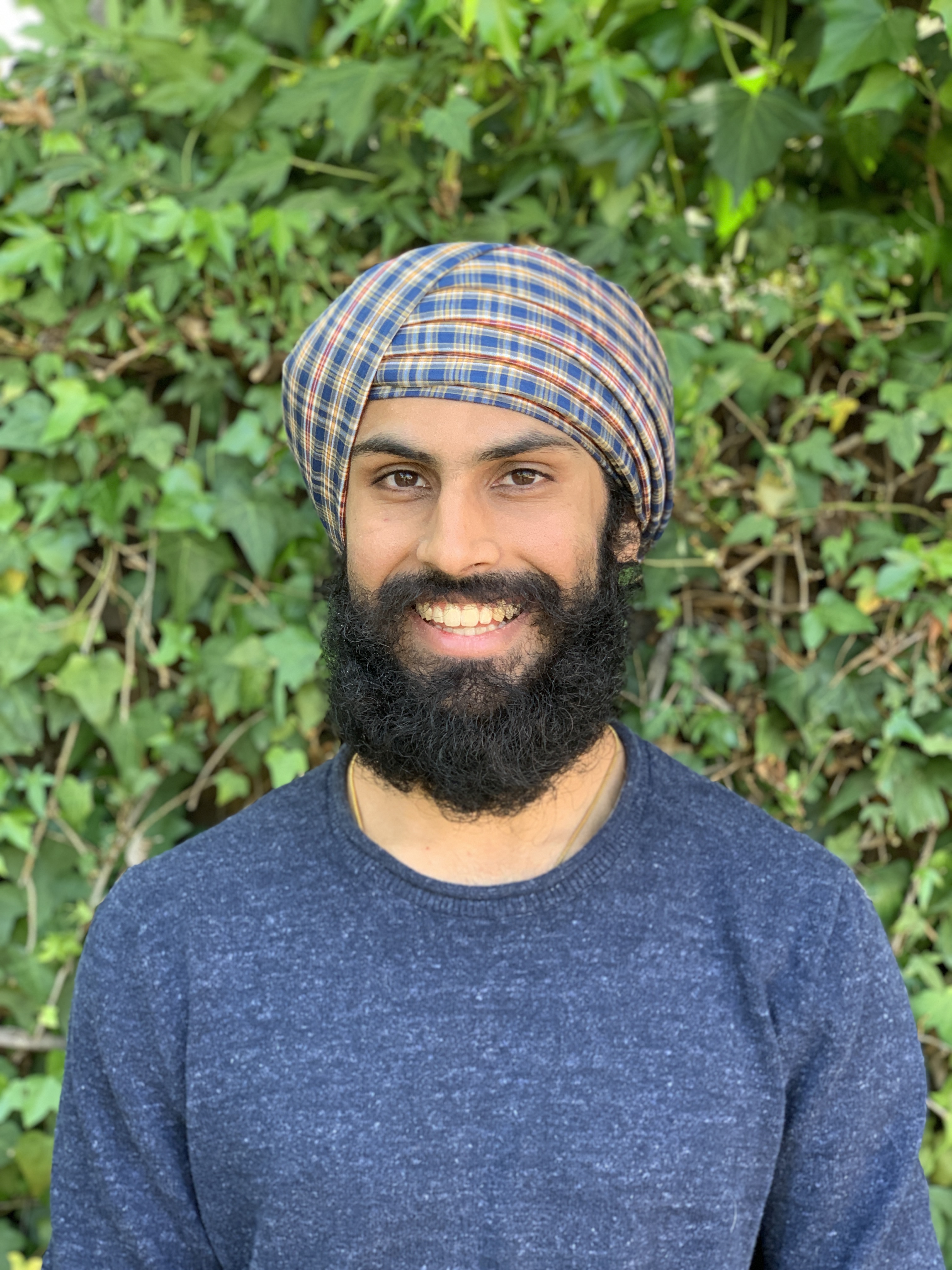}}]{Amit Pal Singh Kohli} is a 5th year Ph.D. student in Laura Waller's group at UC Berkeley. He did his B.S. in Electrical Engineering at Stanford University prior to coming to Berkeley. He works at the intersection of optics, statistics, and machine learning to build robust computational imaging systems for biomedical applications.

\end{IEEEbiography}

\begin{IEEEbiography}[{\includegraphics[width=1in,height=1in,clip,keepaspectratio]{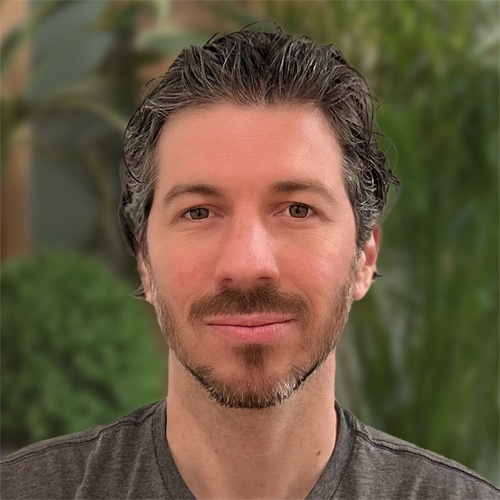}}]{Jacob L. Yates} is an Assistant Professor in the Herbert Wertheim School of Optometry and Vision Science at UC Berkeley. He is affiliated with the Helen Wills Neuroscience Institute and the Redwood Center for Theoretical Neuroscience. He received his PhD from UT Austin in 2016 and was a postdoctoral researcher at the University of Rochester and the University of Maryland, College Park. 

\end{IEEEbiography}

\begin{IEEEbiography}[{\includegraphics[width=1in,height=1in,clip,keepaspectratio]{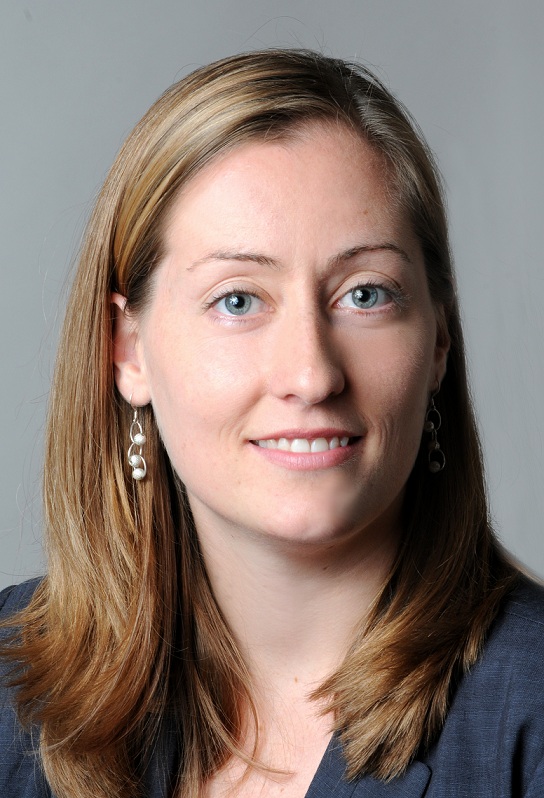}}]{Laura Waller} is the Charles A. Desoer Professor of Electrical Engineering and Computer Sciences at UC Berkeley. She received BS, MEng and PhD degrees from the Massachusetts Institute of Technology in 2004, 2005 and 2010. After that, she was a Postdoctoral Researcher and Lecturer of Physics at Princeton University from 2010-2012. She is a Packard Fellow for Science \& Engineering, OSA Fellow, and Chan-Zuckerberg Biohub Investigator.

\end{IEEEbiography}




\fi

\clearpage                

\begin{center}
  {\LARGE\bfseries Supplementary Material}
\end{center}
\vspace{2em}

\setcounter{section}{0}
\renewcommand\thesection{S\arabic{section}}
\renewcommand\thesubsection{\thesection.\arabic{subsection}}
\setcounter{figure}{0}             
\renewcommand\thefigure{S\arabic{figure}}  
\setcounter{table}{0}
\renewcommand\thetable{S\arabic{table}}



\IEEEraisesectionheading{
  \section{Additional Experimental Results}\label{sec:introduction}
}
\subsection{Motion Dynamic Range}
%
%
%
%
\IEEEPARstart{T}{he} motion dynamic range is important when capturing subtle vibrations. Unlike conventional cameras, event cameras have configurable motion sensitivity via a custom contrast threshold, such that weak light changes are filtered out as noise. To evaluate the system’s response across different motion ranges, we conducted an experiment in which the system’s behavior was assessed as a function of shift size (which we approximated as the input audio amplitude). 

Specifically, we played an up-chirp signal with amplitude ranging from 0.1 to 1.0 in increments of 0.1. To measure how the system's performance degrades, we compute the maximum reconstructible frequency for the default threshold value. Before the experiment, we reduced the speaker volume to calibrate the system such that the maximum reconstructible frequency was approximately 2.3 kHz when playing the chirp signal at an amplitude of 1.0.

\begin{figure}[!t]
\centering
\includegraphics[width=0.4 \textwidth]{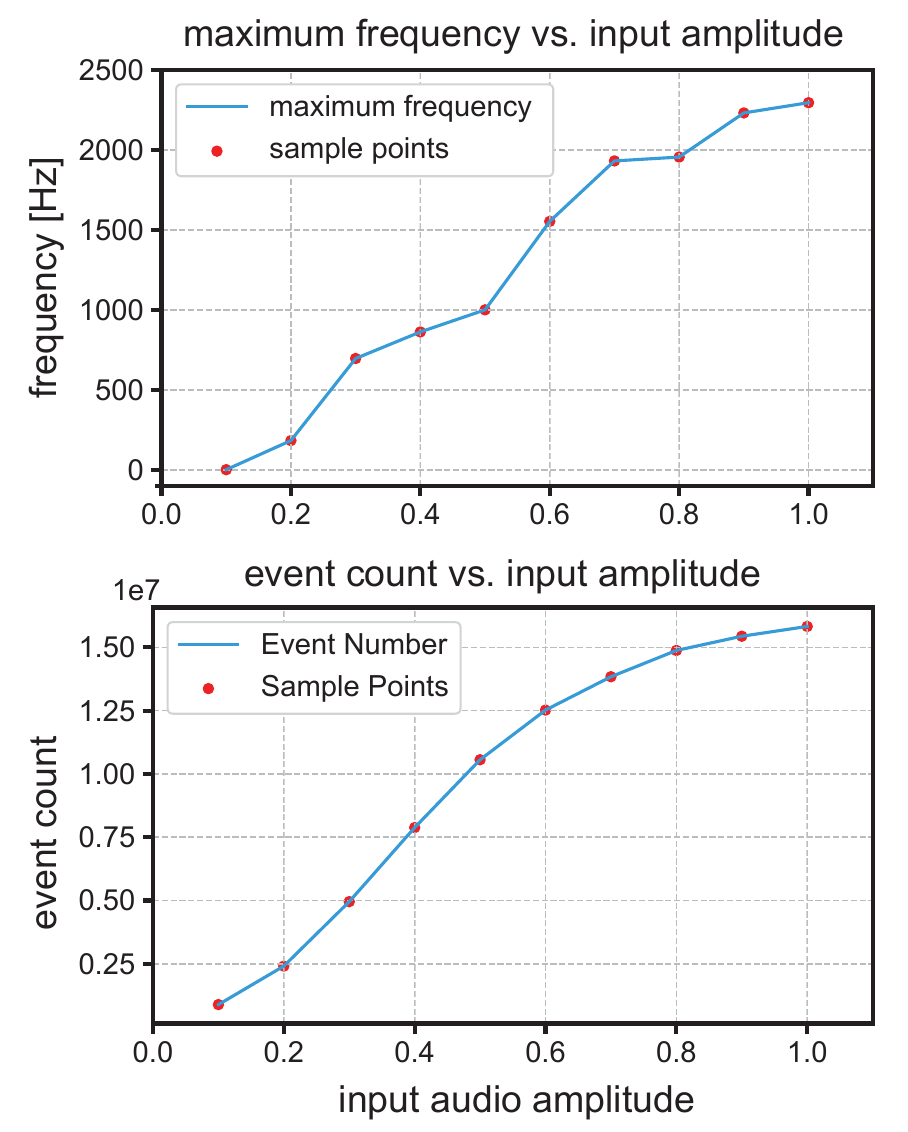}
\caption{Experimental results of motion dynamic range evaluation. \textbf{(Top)} The maximum reconstructible frequency with different input audio amplitudes from 0.1 to 1.0. \textbf{(Bottom)} The recorded event count with different input audio amplitudes from 0.1 to 1.0.}
\label{fig:motionrange}
\end{figure}

In Fig. \ref{fig:motionrange}(top), the maximum reconstructible frequency decreases progressively with lower amplitudes (e.g., 1954 Hz at amplitude 0.8; 695 Hz at amplitude 0.3). At an amplitude of 0.1, no discernible signal was observed in the reconstruction, indicating a failure to capture the vibration under this condition. In parallel, we computed the number of events recorded at each amplitude level in Fig. \ref{fig:motionrange}(bottom), which exhibited a logarithmic growth trend from 0.1 to 1.0. Notably, the trend of the maximum reconstructible frequency roughly followed that of the event count.

Therefore, a sufficient large motion or carefully adjusted motion threshold is needed for high-quality event-based vibration sensing. In our study, our imaging system has the ability to amplify subtle motion by adjusting the focus, which may help mitigate this issue.

\subsection{Large Motion}

\begin{figure*}[!h]
\centering
\includegraphics[width=0.8\textwidth]{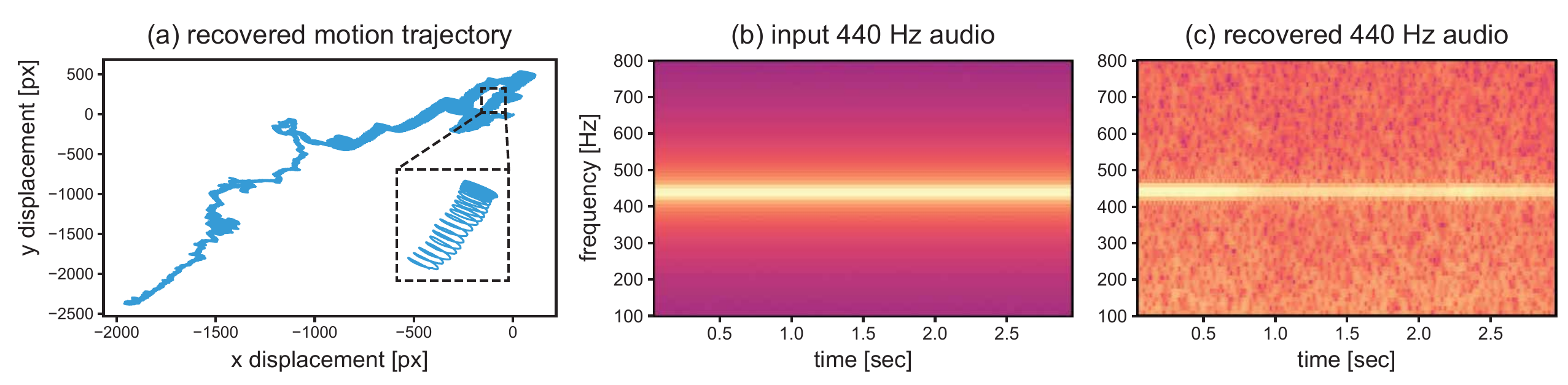}
\caption{Experiment results with large hand-holding motion. \textbf{(a)} Recovered motion trajectory. \textbf{(b)} Input 440 Hz audio. \textbf{(c)} Recovered 440 Hz audio.}
\label{fig:largemotion}
\end{figure*}

We conducted a proof-of-concept experiment to evaluate the system's performance under large motion. We played a 440 Hz signal (Fig. \ref{fig:largemotion}(b)) while hand-holding the speaker to induce large scene motion. The induced motion displacement reaches $\sim$2000 pixels in the X dimension and $\sim$3000 pixels in the Y dimension (Fig. \ref{fig:largemotion}(a)), which is comparable to the large motion reported in \cite{sheinin2022dual}.

When zooming in on the motion trajectory in Fig. \ref{fig:largemotion}(a), the waveform corresponding to the 440 Hz signal becomes clearly visible. Consequently, in the reconstruction shown in Fig. \ref{fig:largemotion}(c), we successfully recovered the 440 Hz audio despite the presence of substantial scene motion.

\begin{figure*}[!h]
\centering
\includegraphics[width=1.0\textwidth]{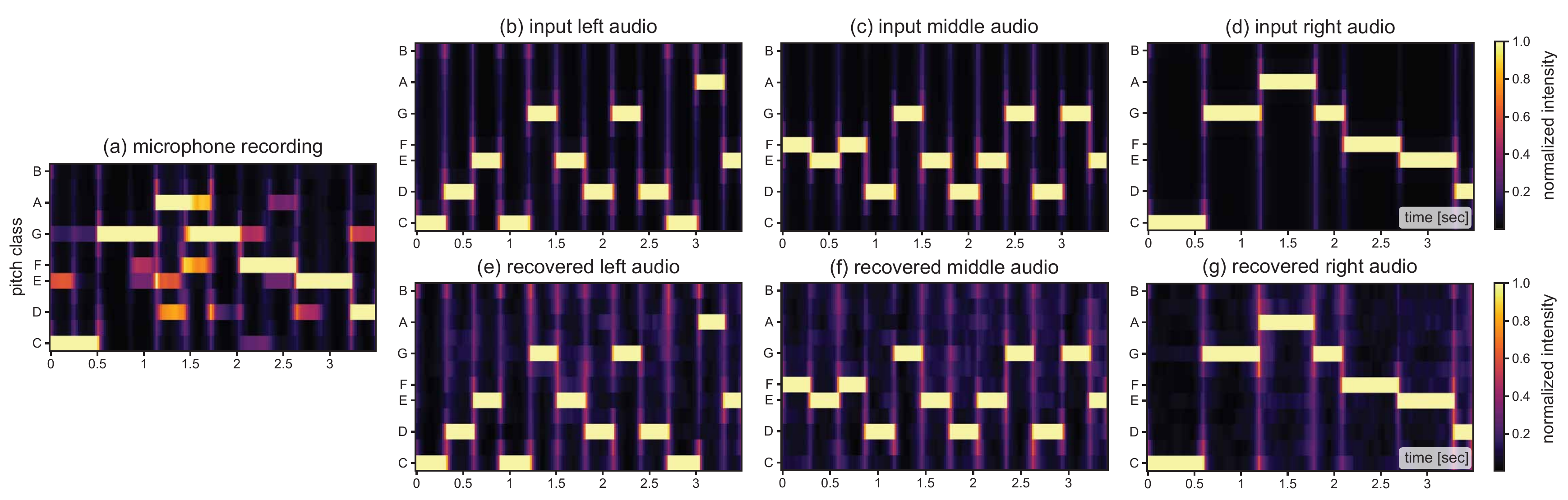}
\caption{Capturing signals from three laser spots and audio sources. \textbf{(a)} The chromagram of the microphone recording. \textbf{(b-d)} The chromagrams of the input audio from left, middle, and right speakers. \textbf{(e-g)} The chromagrams of the recovered left, middle, and right audio signals.}
\label{fig:3laserspots}
\end{figure*}

\subsection{Multiple Laser Spots}
We conducted an additional experiment involving more than two laser spots and audio sources. Specifically, we implemented our system with three laser spots and corresponding audio sources. Consistent with our two-source experiment, speckle patterns from different sources were projected onto distinct regions of the sensor, enabling physical separation of the signals.

In this setup, we played different tones corresponding to different pitch classes through three separate speakers (Figs. \ref{fig:3laserspots}(b–d)) while simultaneously recording the mixed audio with a microphone. As shown in Fig. \ref{fig:3laserspots}(a), the microphone captured a complex superposition of the three audio signals. In contrast, our system successfully isolated and reconstructed the individual audio signals, as shown by the accurate and clearly resolved chromagrams in Figs. \ref{fig:3laserspots}(e–g).

Other than audio source separation, Zhang \textit{et al.} \cite{Zhang2023MultiSpot} demonstrated that having more laser spots enables the analysis of surface vibration waves. This in turned allowed for impact localization and analysis of material properties, which are exciting avenues for future exploration of event-based vibrometry.

\end{document}